\PassOptionsToPackage{prologue,dvipsnames}{xcolor}
\documentclass[aps,prf,onecolumn,showpacs,groupedaddress,superscriptaddress,footinbib,longbibliography]{revtex4-2}
\usepackage{natbib}
\usepackage{mathrsfs}
\usepackage{hyperref}
\setcitestyle{square,sort&compress,comma,numbers}
\hypersetup{colorlinks=true, citecolor=blue, urlcolor=blue, linkcolor=blue}
\usepackage[english]{babel}
\usepackage[utf8]{inputenc}
\usepackage{graphicx}
\usepackage{amsmath}
\usepackage{amsfonts}
\usepackage{amssymb}
\usepackage{color,soul}
\setulcolor{red}
\usepackage{easyReview}
\usepackage{xcolor}
\usepackage{booktabs}

\usepackage[normalem]{ulem}

\begin{document}

\title{Impact of Taylor--Aris dispersion on chemical patterns in confined channel flows}
\author{Yanis Baouche}
\affiliation{Max-Planck-Institut f{\"u}r Physik komplexer Systeme, N{\"o}thnitzer Stra{\ss}e 38,
01187 Dresden, Germany}
\author{Ramita Mondal}
\affiliation{Max-Planck-Institut f{\"u}r Physik komplexer Systeme, N{\"o}thnitzer Stra{\ss}e 38,
01187 Dresden, Germany}
\affiliation{School of Mathematics, Fry Building, University of Bristol, BS8 1UG, UK}
\author{Christina Kurzthaler}
\email{ckurzthaler@pks.mpg.de}
\affiliation{Max-Planck-Institut f{\"u}r Physik komplexer Systeme, N{\"o}thnitzer Stra{\ss}e 38,
01187 Dresden, Germany}
\affiliation{Center for Systems Biology Dresden, Pfotenhauerstra{\ss}e 108, 01307 Dresden, Germany}
\affiliation{Cluster of Excellence, Physics of Life, TU Dresden, Arnoldstra{\ss}e 18, 01062 Dresden, Germany}

\begin{abstract} 
Chemical patterns form through non-linear interactions between different species. We study how confined channel flows affect pattern formation using the Brusselator as a paradigmatic activator--inhibitor model. An invariant-manifold method for slender channels reduces the advection--diffusion--reaction equation to one spatial dimension, with species transported at the mean flow velocity and molecular diffusivities replaced by the classical Taylor--Aris dispersion coefficients. We obtain the phase diagram as a function of the Péclet number (i.e., the relative strength of advection to diffusion), the channel slenderness, and the diffusivity ratio of the species. Our analytical results demonstrate that the classical Turing criterion can become inverted at large P{\'e}clet numbers and, thus,  channel flow can trigger pattern formation even when the molecular diffusivities violate the classical Turing condition, and conversely destroy pre-existing patterns. Both predictions are confirmed by direct numerical simulations, which show advected stripe patterns, whose wavelength can be predicted by our reduced theory, as well as oscillatory dynamics. These findings highlight that hydrodynamic flows can modify chemical mixing and thereby profoundly affect pattern formation in channel geometries.
\end{abstract}

\maketitle

\section{Introduction}
Pattern formation is a central theme in the study of biological systems~\cite{kochBiologicalPatternFormation1994a, ermentroutMathematicalTheoryVisual1979, murrayMathematicalBiology1989} and has received substantial attention in various other contexts, including  chemical oscillators~\cite{tysonWhatEveryoneShould1994, prigogineSymmetryBreakingInstabilities1968} and interfacial instabilities~\cite{crossPatternFormationOutside1993}. The mechanisms behind pattern formation are diverse: while activator--inhibitor dynamics can induce Turing patterns in chemical species~\cite{kochBiologicalPatternFormation1994a}, mechano-chemical couplings in active systems~\cite{osterMechanicalAspectsMesenchymal1983,mietkeMinimalModelCellular2019a} or couplings with the environment, including binding to and unbinding from   boundaries~\cite{otsujiMassConservedReaction2007,goryachevDynamicsCdc42Network2008,moriWavePinningCellPolarity2008}, can also give rise to intriguing patterns with profound implications for biological systems. Additionally, patterns have been engineered in synthetic settings, from gene circuits in bacterial populations to in vitro reconstitution of protein systems~\cite{basuSyntheticMulticellularSystem2005,looseSpatialRegulatorsBacterial2008}, offering controlled platforms to test theoretical predictions.
In realistic systems, however, patterns often form in confined and moving fluids, which advect interacting species and can thereby alter the formation of spatial structure.  Developmental biology offers prime examples: morphogens responsible for establishing left--right asymmetry in the vertebrate embryo are transported by cilia-driven flow of extraembryonic fluid~\cite{nonakaRandomizationLeftRight1998,hirokawaNodalFlowGeneration2006}, while cytoplasmic streaming in elongated cells provides a robust transport mechanism for proteins and organelles
across cellular length scales, affecting intracellular organization and mixing~\cite{goldsteinPhysicalPerspectiveCytoplasmic2015,quinlanCytoplasmicStreamingDrosophila2016,htetAnalyticalMethodsCytoplasmic2025}. Understanding the role of flows in pattern formation may thus not only provide crucial insights into cell biology but also have fundamental implications for biochemistry, where chemical reactions typically take place in aqueous solution~\cite{MolecularBiologyCell2015} and flows may represent a route for engineering new spatio-temporal patterns.

Despite the importance of this question, few analytical predictions have been made for the impact of flow on pattern formation. A mechanistically distinct route is the differential-flow-induced chemical instability, where the activator and inhibitor are advected at different rates~\cite{rovinskyChemicalInstabilityInduced1992,rovinskySelforganizationInducedDifferential1993,rovinskyDifferentialFlowInstability1994,balinskykhazanDifferentialFlowInduced1998,stucchiPatternFormationInduced2013,stucchiPatternFormationInduced2014}. In addition, a linear theory for reaction--diffusion--advection instabilities in Stokes flow across several channel geometries was developed in~\cite{vangorderDiffusiveInstabilitiesSpatial2019}. By
averaging the flow over the channel cross-section, their theory captures the resulting spatio-temporal character of the pattern, but the Turing threshold itself remains unchanged under
equal advection of both species. Closest to our work, Vasquez~\cite{vasquezChemicalInstabilityInduced2004} showed through a discretized linear stability analysis of the two-dimensional Brusselator together with a
confirming simulation that a sufficiently strong Poiseuille flow can destabilize an otherwise stable homogeneous state, even though both species share a single velocity field and differ only in their molecular diffusivities. A quantitative theory of when such a flow creates, or destroys, a pattern has so far been missing.

In this work we study the role of confined fluid flows in pattern formation. Using an invariant-manifold reduction, we derive effective one-dimensional equations for a two-species Brusselator reaction--diffusion--advection system in a slender planar channel with Poiseuille flow. We find that the molecular diffusivities become replaced by the shear-induced Taylor-Aris diffusivities, which modify the classical Turing threshold. Importantly, our analytical results show that for strong flows the Turing criterion becomes inverted. As a consequence,  channel flow can both trigger a finite-wavelength instability when the two species' diffusivities violate the classical Turing ordering, and suppress one when that ordering is satisfied. Our model yields three explicit thresholds, two of which depend on the flow, separating stable, pattern-forming, competing, and uniformly oscillating regimes. We compare them to the two-dimensional linear stability analysis and to full numerical simulations. The latter reveal stripes that drift with the mean flow and bend with the Poiseuille profile. The wavelength of the stripes increases with the P{\'e}clet number, matching the predictions of our reduced theory. Our non-linear simulations further demonstrate that stripe patterns win the competition against uniform oscillations well beyond the region predicted by linear theory.

\section{Model setup \label{sec:model}}
\begin{figure}[htp]
    \centering
    \includegraphics[width=\columnwidth]{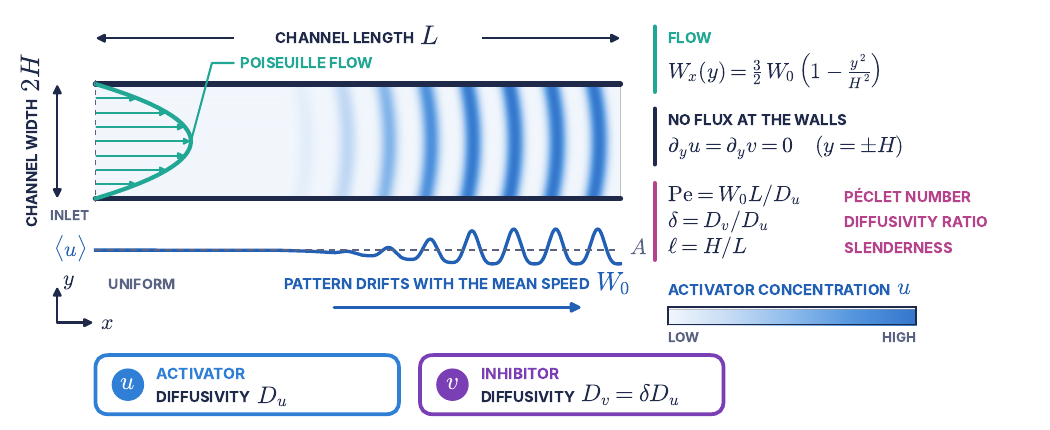}
    \caption{Model setup. Two species, an activator $u$ and an inhibitor $v$, diffuse with molecular diffusivities $D_u$ and $D_v$. They react via Brusselator kinetics and are advected by a planar Poiseuille flow $\boldsymbol{W}=W_x(y)\boldsymbol{\hat x}$ in a channel of length $L$ and width~$2H$.}
    \label{fig:schema11}
\end{figure}

We consider a system of chemical species with concentrations $\boldsymbol{c}(\boldsymbol{r},t)=[c_1(\boldsymbol{r},t), \dots, c_N(\boldsymbol{r},t)]$ that depend on space $\boldsymbol{r}=(x,y)$ and time $t$. The species react with each other, diffuse, and are advected by an incompressible fluid flow  (see Fig.~\ref{fig:schema11} for a two-species example). Their dynamics obey the advection--diffusion--reaction equation:
\begin{align}
\frac{D \boldsymbol{c}}{Dt}:= \partial_t \boldsymbol{c}+\boldsymbol{W}\cdot\boldsymbol{\nabla }\boldsymbol{c} = \boldsymbol{D}\nabla^2\boldsymbol{c}+\boldsymbol{R}.
\end{align}
Here, $\boldsymbol{D}=\mathrm{diag}(D_1, \dots, D_N)$ denotes the constant diffusion matrix, $\boldsymbol{W}$ is the fluid velocity, and $\boldsymbol{R}=\boldsymbol{R}(\boldsymbol{c})$ is the reaction vector, which includes non-linear couplings among the different chemical species. The chemical species are immersed in a two-dimensional channel of length $L$ and width $2H$: $0\leq x\leq L$ and $-H\leq y\leq H$. We assume a pressure-driven flow along the channel having a Poiseuille flow profile:  
\begin{align}
\boldsymbol{W} &= W_x(y) \hat{\boldsymbol{x}} \quad \text{with}\quad W_x(y) = \frac{3}{2} W_0 \left( 1 - \frac{y^2}{H^2}\right),
\end{align}
where $W_0$ is the cross-sectionally averaged velocity. The latter is defined as
\begin{align}
W_0 := \bar{W}_x=\frac{1}{2H}\int_{-H}^H W_x(y)~\mathrm{d}y.
\end{align}
Throughout, we denote the cross-sectional average of a quantity $f$ by an overbar or by angle brackets $\bar{f} = \langle f \rangle$, using the bar for single symbols and the brackets for products and longer expressions.
We further impose no-flux boundary conditions for the chemical fields $\partial_y\boldsymbol{c}|_{y=\pm H}=\boldsymbol{0}$ at the boundaries and periodic boundary conditions at the in- and outlets.  

\subsection{Non-dimensional quantities}
We consider two characteristic time scales of the system: first, the time it takes a chemical species $c_i$ to diffuse across the  half-width $\tau_{{\rm diff},i}=H^2/D_i$ and second, the time to traverse the channel by advection $\tau_{\rm adv}=L/W_0$, giving a non-dimensional parameter 
\begin{align}
\epsilon_i = \frac{\tau_{{\rm diff},i}}{\tau_{\rm adv}} 
= \frac{W_0L}{D_i}\frac{H^2}{L^2} := {\rm{Pe}}_i \ell^2.
\end{align}
Here, $\ell=H/L$ denotes the aspect ratio (or slenderness) of the channel and the Péclet number  characterizes the relative strength of advection along the channel with respect to diffusion of species $i$, $\mathrm{Pe}_i = W_0 L/D_i$. We can further write the non-dimensional parameter for other species as $\epsilon_j=\epsilon_i/\delta_{ij}$, which depends on the ratio of molecular diffusivities  
\begin{align}
\delta_{ij} = \frac{D_j}{D_i}.
\end{align}
Lastly, we introduce the characteristic time of the chemical reaction kinetics $\tau_{\rm chem}$. This provides another non-dimensional parameter, the Damköhler number, which measures how fast the reactions are compared with advection along the channel, $\mathrm{Da}=\tau_\mathrm{adv}/\tau_\mathrm{chem}$.  Compared
with transverse diffusion, the reactions are slow when
$\tau_{\mathrm{diff},i}/\tau_\mathrm{chem}=\epsilon_i\,\mathrm{Da}\ll1$. Unless stated otherwise, we use $\mathrm{Da}=1$.

\subsection{Derivation of an effective advection--diffusion--reaction equation in a long slender channel}
To make analytical progress, we consider a long slender channel with a small aspect ratio, $\ell\ll 1$. We assume that the diffusion across the channel is much faster than advection along its long axis, $\epsilon_i\ll1$ for $i=1,\dots N$, allowing for a separation of scales. Since $\epsilon_{j} = \epsilon_i / \delta_{ij}$, this condition is most restrictive for the slowest-diffusing species. 
This allows us to perform an invariant-manifold reduction, thus reducing our spatially two-dimensional problem to a one-dimensional one. Therefore, we rescale the concentration fields by a constant base concentration $\boldsymbol{c}'=\boldsymbol{c}/c_0$, length scales by $x'=x/L$ and $y'=y/H$, the time scale by $t'=t/\tau_{\rm adv}$,  and the velocity by $\boldsymbol{W}' = \boldsymbol{W}/W_0 = W'_{x'}(y')\hat{\boldsymbol{x}}$ so that $W'_{x'}(y') = \frac{3}{2}\left(1-y'^2 \right)$ has unit cross-sectional average.
The reaction term is rescaled by $\boldsymbol{R}'=\boldsymbol{R}/(c_0/\tau_{\rm chem})$, leading to the non-dimensional reaction--diffusion--advection equation: 
\begin{align}
\partial_{t'}\boldsymbol{c}'+\boldsymbol{W}'\cdot \boldsymbol{\nabla}'\boldsymbol{c}' &= \frac{\boldsymbol{D}'}{{\rm Pe}_1}\partial_{x'}^2\boldsymbol{c}'+ \frac{\boldsymbol{D}'}{\epsilon_1}\partial_{y'}^2\boldsymbol{c}' + {\rm Da}\boldsymbol{R}',
\end{align}
with $\boldsymbol{D}' = \mathrm{diag}(1,\delta_{12},\dots,  \delta_{1N})$. We further rewrite  $1/{\rm Pe}_1= \tau_{\rm adv}/\tau_{{\rm diff},1}^L = \epsilon_1\beta$ with $\tau_{{\rm diff},1}^L=L^2/D_1$ and $\beta=\mathcal{O}(1)$, keeping molecular diffusion along the long axis of the channel slower than advection. 
The modified equation then reads
\begin{align}
\partial_{t'}\boldsymbol{c}'+\boldsymbol{W}'\cdot \boldsymbol{\nabla}' \boldsymbol{c}' &= \epsilon_1 \beta \, \boldsymbol{D}' \partial_{x'}^2\boldsymbol{c}'+ \frac{\boldsymbol{D}'}{\epsilon_1}\partial_{y'}^2\boldsymbol{c}' + {\rm Da}\boldsymbol{R}',\label{eq:c_nondim}
\end{align}
where we assume the reactions to be slow compared to transverse diffusion, $\epsilon_i\mathrm{Da}\ll1$ for all species. For $\mathrm{Da}=\mathcal{O}(1)$ this reduces to $\epsilon_1 \ll 1$. We now seek a reduced perturbative equation in $\epsilon_1\ll 1$ of the form~\cite{mercerCentreManifoldDescription1990,marbachActiveControlDispersion2019}: 
\begin{align}
\partial_{t'}\bar{\boldsymbol{c}}' = \boldsymbol{G}^{(1)}[\bar{\boldsymbol{c}}', \partial_{x'} \bar{\boldsymbol{c}}']+ \epsilon_1 \boldsymbol{G}^{(2)}[\bar{\boldsymbol{c}}', \partial_{x'} \bar{\boldsymbol{c}}', \partial_{x'}^2\bar{\boldsymbol{c}}']+\mathcal{O}(\epsilon_1^2),
\end{align}
where the cross-sectionally averaged concentration field is $\bar{\boldsymbol{c}}'\equiv\bar{\boldsymbol{c}}'(x', t')$ and the concentration field is expanded as
\begin{align}
\boldsymbol{c}' = &\boldsymbol{V}^{(0)}[y',t'; \bar{\boldsymbol{c}}']+ \epsilon_1 \boldsymbol{V}^{(1)}[y',t';\bar{\boldsymbol{c}}', \partial_{x'} \bar{\boldsymbol{c}}']+ \epsilon_1^2 \boldsymbol{V}^{(2)}[y',t'; \bar{\boldsymbol{c}}', \partial_{x'} \bar{\boldsymbol{c}}', \partial_{x'}^2\bar{\boldsymbol{c}}']+\mathcal{O}(\epsilon_1^3). \label{eq:expansion_c_prime}
\end{align}
Expanding the reaction term gives
\begin{align}
\boldsymbol{R}'(\boldsymbol{c}')&= \boldsymbol{R}'(\boldsymbol{V}^{(0)})+\epsilon_1 \boldsymbol{J}(\boldsymbol{V}^{(0)})\boldsymbol{V}^{(1)} +\mathcal{O}(\epsilon^2_1),
\end{align}
where $\boldsymbol{J}$ denotes the Jacobian.  Multiplying Eq.~\eqref{eq:c_nondim} by $\epsilon_1$ and inserting these expansions, we arrive at a set of equations for each order in $\epsilon_1$: 
\begin{subequations}
\begin{align}
&\epsilon_1^0: \,  \boldsymbol{D}' \partial_{y'}^2 \boldsymbol{V}^{(0)}  =\boldsymbol{0},\\
&\epsilon_1: \,   \boldsymbol{D}'\partial_{y'}^2\boldsymbol{V}^{(1)} = \boldsymbol{G}^{(1)} + W_{x'}'\partial_{x'}\bar{\boldsymbol{c}}'- \mathrm{Da}\,\boldsymbol{R}'({\boldsymbol{V}}^{(0)}), \label{eq:eps1}\\
\begin{split}
&\epsilon_1^2: \,  \boldsymbol{D}'\partial_{y'}^2\boldsymbol{V}^{(2)} = 
\boldsymbol{G}^{(2)} + \partial_{t'}\boldsymbol{V}^{(1)} + W_{x'}'\partial_{x'}\boldsymbol{V}^{(1)}  - \beta \boldsymbol{D}'\partial_{x'}^2\bar{\boldsymbol{c}}'  - \mathrm{Da}\,\boldsymbol{J}(\boldsymbol{V}^{(0)})\boldsymbol{V}^{(1)}.
\end{split}
\end{align}
\end{subequations}
The no-flux boundary conditions translate into no-flux conditions for each order $\boldsymbol{V}^{(i)}$. The cross-sectional average gives $\langle{\boldsymbol{V}}^{(0)}\rangle=\bar{\boldsymbol{c}}'$, showing that the leading-order concentration is constant across the channel. Since $\bar{\boldsymbol{c}}'$ is by definition the cross-sectional average of $\boldsymbol{c}'$, averaging Eq.~\eqref{eq:expansion_c_prime} requires $\langle{\boldsymbol{V}}^{(i)}\rangle =\boldsymbol{0}$ for $i \geq 1$, which closes the system. Averaging the first-order equation [Eq.~\eqref{eq:eps1}] provides the relation
\begin{align}
\boldsymbol{G}^{(1)} &=  -\partial_{x'} \bar{\boldsymbol{c}}' + \mathrm{Da}\, \boldsymbol{R}'(\bar{\boldsymbol{c}}'), \label{eq:G1}
\end{align}
as $\bar{W}'_{x'}=1$ and due to the no-flux boundary conditions the cross-sectional average $\langle \partial_{y'}^2 \boldsymbol{V}^{(1)}\rangle$ vanishes. Similarly, we obtain for the second order 
\begin{align}
\boldsymbol{G}^{(2)} = \beta \boldsymbol{D}'\partial_{x'}^2\bar{\boldsymbol{c}}' - \langle W_{x'}'\partial_{x'}\boldsymbol{V}^{(1)}\rangle, 
\end{align}
which requires knowledge of the explicit form of $\boldsymbol{V}^{(1)}$. The latter can be obtained by inserting Eq.~\eqref{eq:G1} into Eq.~\eqref{eq:eps1}, leading to 
\begin{align}
\boldsymbol{V}^{(1)} &=  \left(\frac{y'^2}{4} - \frac{y'^4}{8} - \frac{7}{120}\right)\,(\boldsymbol{D}')^{-1}\partial_{x'}\bar{\boldsymbol{c}}'.
\end{align}
Using this as input, the reduced equation becomes: 
\begin{align}
\partial_{t'} \bar{\boldsymbol{c}}'+ \partial_{x'}\bar{\boldsymbol{c}}' = \boldsymbol{D}_{\rm eff}'\partial_{x'}^2\bar{\boldsymbol{c}}'+\boldsymbol{R}'_{\rm eff} +\mathcal{O}(\epsilon_1^2),
\end{align}
with effective diffusion tensor and reaction vector: 
\begin{align}
D'_{{\rm eff}, i}= \frac{1}{{\rm Pe}_i} +\frac{2}{105}\epsilon_i  \qquad \text{and} \qquad \boldsymbol{R}'_{\rm eff}= {\rm Da}\boldsymbol{R}'(\bar{\boldsymbol{c}'}).\label{eq:eff_diff}
\end{align}
Moving to the co-moving frame of reference by substituting $\xi=x'-t'$, we arrive at an effective reaction--diffusion equation:
\begin{align}
\partial_{t'} \bar{\boldsymbol{c}}' = \boldsymbol{D}'_{\rm eff}\partial_{\xi}^2\bar{\boldsymbol{c}}'+\boldsymbol{R}'_{\rm eff}. \label{eq:reduced_effective}
\end{align}
It is important to note that the flow enters the reduced dynamics only through the shear-induced effective diffusivity $\boldsymbol{D}'_{\rm eff}$ of the chemical species. This phenomenon, known as the classical Taylor--Aris dispersion~\cite{taylorDispersionSolubleMatter1953,arisDispersionSoluteFluid1956}, results from the fact that the solutes at the center of the channel experience a different fluid velocity than those close to the boundaries. Due to molecular diffusion across the channel (and thus `jumps' across streamlines), the dispersion of solute along the channel axis becomes enhanced. In dimensional form, the effective diffusivity reads: 
\begin{align}
D_{{\rm eff}, i}= D_i +\frac{2}{105}\frac{W_0^2H^2}{D_i}, \label{eq:eff_diff_dim}
\end{align}
showing that a Poiseuille flow in a planar channel enhances the molecular diffusivities of the individual species as $D_{\rm eff,i}\sim  H^2W_0^2/D_i$. This effectively changes the mixing of chemical species along the channel, which can amplify pattern formation. Importantly, shear affects dispersion of slower-diffusing species more than that of faster-diffusing ones. To predict the latter we use the reduced equation [Eq.~\eqref{eq:reduced_effective}] as the starting point for the stability analysis of the non-linear reaction kinetics.

\section{Linear stability analysis in a slender channel}

The reduced equation~\eqref{eq:reduced_effective} is a reaction--diffusion system in which the flow enters only through the effective diffusivities, and its linear stability therefore depends on the kinetics only through the reaction Jacobian. In this section, we perform a stability analysis for a general two-species reaction. To this end, we first simplify the notation and denote: $u=\bar{c}_1'$, $v=\bar{c}_2'$, $d_u=D_{\mathrm{eff},1}'$, $d_v=D_{\mathrm{eff},2}'$,  $\mathrm{Pe}=\mathrm{Pe}_1$, $\epsilon=\epsilon_1=\mathrm{Pe} \ell^2$, and $\delta=\delta_{12}=D_v/D_u$.

\subsection{General two-species kinetics}
\label{sec:lsa-general}
Writing $\boldsymbol{R}'=(f,g)^\top$, the reduced dynamics read
\begin{subequations}\label{eq:red}
\begin{align}
\partial_t u &= d_u\partial_\xi^2 u
  + \mathrm{Da} f(u,v), \label{eq:red-u}\\
\partial_t v &= d_v\partial_\xi^2 v
  + \mathrm{Da} g(u,v), \label{eq:red-v}
\end{align}
\end{subequations}
with effective diffusivities [Eq.~\eqref{eq:eff_diff}]
\begin{equation}
    d_u= \frac{1}{\mathrm{Pe}}+\frac{2}{105}\epsilon \quad\text{and}\quad
    d_v=\frac{\delta}{\mathrm{Pe}}
  +\frac{2}{105} \frac{\epsilon}{\delta}.
\label{eq:dudv}
\end{equation}
We first denote by $(u^\star,v^\star)$ the stationary, homogeneous state where $f(u^\star,v^\star)=g(u^\star,v^\star)=0$, linearize
Eqs.~\eqref{eq:red} around it, and look for perturbations
$\delta \boldsymbol{w}(\xi,t)=\hat{\boldsymbol{w}} e^{\lambda t+iq\xi}$, with dimensionless axial wavenumber $q$ and growth rate $\lambda(q)$, and $\delta \boldsymbol{w}=(\delta u,\delta v)^\top$. This gives the
eigenvalue problem $\lambda \delta \boldsymbol{w}=\boldsymbol{J}_q \delta \boldsymbol{w}$
with
\begin{equation}
    \boldsymbol{J}_q=\mathrm{Da} \boldsymbol{J} -q^2 \begin{pmatrix} d_u & 0\\ 
                    0 & d_v
    \end{pmatrix}, \qquad \boldsymbol{J}=\begin{pmatrix} a_{11} & a_{12}\\
  a_{21} & a_{22}\end{pmatrix}
=\left. \frac{\partial(f,g)}{\partial(u,v)}
  \right|_{(u^\star,v^\star)} .
\label{eq:Jq-gen}
\end{equation}
Here, $\boldsymbol{J}$ is the Jacobian of the kinetics alone
and depends neither on $q$, nor on $\mathrm{Da}$, nor on
the flow, whereas $\boldsymbol{J}_q$ is the Jacobian of the
full linearized problem at wavenumber $q$. In particular,
the uniform mode is governed by $\boldsymbol{J}_0=\mathrm{Da} \boldsymbol{J}$, which coincides with $\boldsymbol{J}$ only for $\mathrm{Da}=1$.

The growth rates are given by $\lambda_\pm(q)= \tfrac{1}{2} \left[
\mathrm{Tr}(\boldsymbol{J}_q) \pm \sqrt{\mathrm{Tr}(\boldsymbol{J}_q)^2
-4\det(\boldsymbol{J}_q)} \right]$, with trace and determinant:
\begin{subequations}\label{eq:trdet-gen}
\begin{align}
    \mathrm{Tr}(\boldsymbol{J}_q) &= \mathrm{Da} T_0-q^2 \Sigma, \label{eq:tr-gen}\\
    \det(\boldsymbol{J}_q) &= \mathrm{Da}^2\Delta_0
  -\mathrm{Da} S q^2+q^4 d_u d_v.
\label{eq:det-gen}
\end{align}
\end{subequations}
Here, we denote the trace and determinant of $\boldsymbol{J}$ by $T_0=a_{11}+a_{22}$ and $\Delta_0=a_{11}a_{22}-a_{12}a_{21}$, respectively, and introduced two quantities:
\begin{equation}
    \Sigma=d_u+d_v \qquad {\rm and} \qquad S=a_{11}d_v+a_{22}d_u .
\label{eq:S-gen}
\end{equation}
We note that the mode with wavenumber $q$ is linearly unstable when $\max \mathrm{Re} [\lambda_\pm(q)]>0$.

\subsubsection{Uniform mode and Hopf threshold}
To study the stability of the uniform mode, we set $q=0$, whereby the diffusion drops out and $\mathrm{Tr}(\boldsymbol{J}_0)=\mathrm{Da} T_0$ and
$\det(\boldsymbol{J}_0)=\mathrm{Da}^2\Delta_0$. So the homogeneous state is stable to uniform perturbations if 
\begin{equation}
    T_0<0 \quad\text{and} \quad \Delta_0>0 .
\label{eq:uniform-stab}
\end{equation}
We now assume $\Delta_0>0$. The state then undergoes
a Hopf bifurcation at $T_0=0$, where $\lambda_\pm(0)=\pm i \mathrm{Da}\sqrt{\Delta_0}$ with frequency $\omega_H=\mathrm{Da}\sqrt{\Delta_0}$. Importantly, the Hopf threshold is independent of the flow and the latter cannot change the local reaction threshold for temporal oscillations.

\subsubsection{Finite-wavelength instability}

\begin{figure}[tp]
    \centering
    \includegraphics[width=\columnwidth]{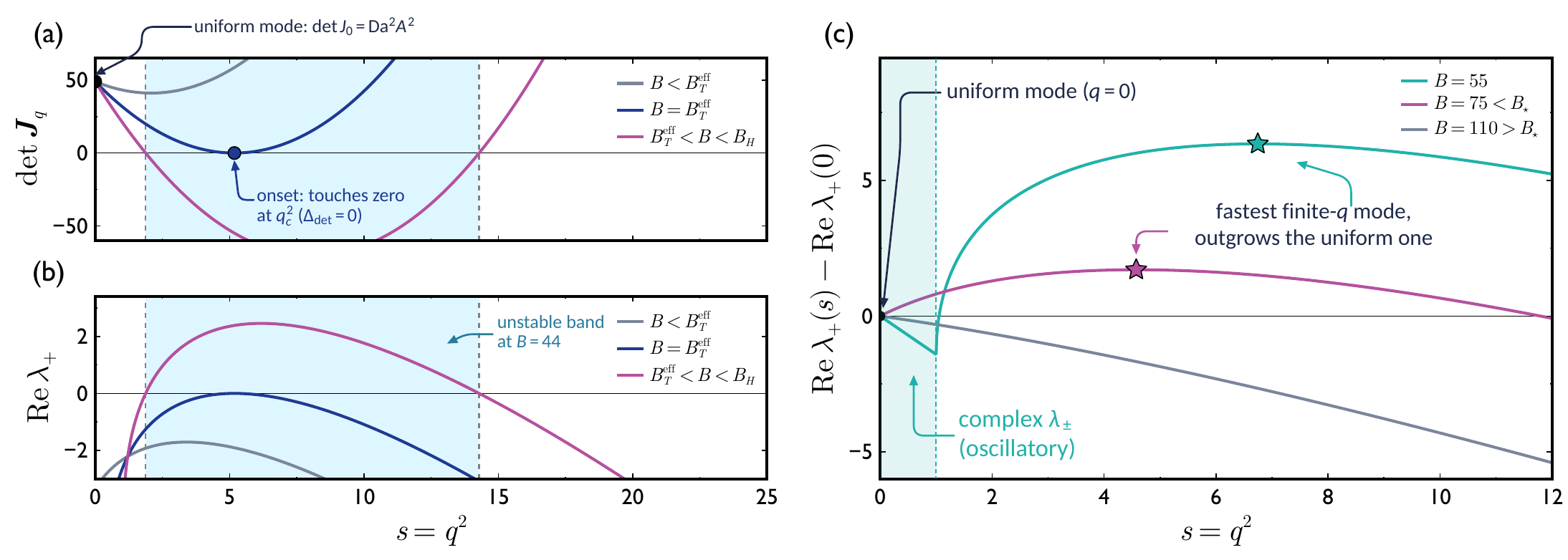}
\caption{Finite-wavelength instability and competition in the reduced theory. (a) Determinant $\det \boldsymbol{J}_q$ [Eq.~\eqref{eq:det-gen}] and (b) growth rate $\mathrm{Re} \lambda_+$ as functions of $s=q^2$, below ($B=32$), at ($B=B_T^\mathrm{eff}=38.2$) and above ($B=44$) the effective Turing threshold; all are below $B_H=50$. Every curve starts at $\det \boldsymbol{J}_0=\mathrm{Da}^2A^2>0$ (black dot). At the onset the parabola touches zero at $q_c^2$ (blue dot, $\Delta_\mathrm{det}=0$). Shaded: the unstable band at $B=44$, where $\det\boldsymbol{J}_q<0$ and $\mathrm{Re}\lambda_+>0$. (c) Above the Hopf threshold $B_H$: growth rate relative to the uniform mode for $B=55$ (inside the window $(A-1)^2<B<(A+1)^2$, where $\lambda_\pm(0)$ are complex), $B=75<B_\star$ and $B=110>B_\star$ ($B_\star=93.7$). Parameters are $A=7$, $\mathrm{Da}=1$, $\mathrm{Pe}\ell=20$, $\delta=0.5$.}
    \label{fig:threshold}
\end{figure}

A stationary finite-wavenumber instability requires the
homogeneous state to remain stable to uniform perturbations
[Eq.~\eqref{eq:uniform-stab}] while one mode $q>0$ becomes
unstable. Since $T_0<0$, the trace
$\mathrm{Tr}(\boldsymbol{J}_q)=\mathrm{Da} T_0-q^2\Sigma$
[Eq.~\eqref{eq:tr-gen}] is negative for all $q$, and the
instability can only come from $\det(\boldsymbol{J}_q)<0$.

As a function of $s=q^2$, Eq.~\eqref{eq:det-gen} is an upward parabola that starts at $\mathrm{Da}^2\Delta_0>0$ [see Fig.~\ref{fig:threshold} (a) for the example of the Brusselator kinetics introduced in Sec.~\ref{sec:brusselator}]. It becomes negative
only if $S>0$ and if its discriminant is positive,
\begin{equation}
    \Delta_{\det}=\mathrm{Da}^2 \left(S^2-4d_u d_v\Delta_0 \right)>0 .
\label{eq:disc}
\end{equation}
Both requirements combine into the instability
condition~\cite{segelDissipativeStructureExplanation1972}
\begin{equation}
    a_{11}d_v + a_{22}d_u > 2\sqrt{d_u d_v \Delta_0}.
\label{eq:inst}
\end{equation}
At the stability onset, Eq.~\eqref{eq:inst} is an equality, the two roots of $\det(\boldsymbol{J}_q)=0$ coincide, and the first unstable wavenumber is 
\begin{equation}
    q_c^2=\frac{\mathrm{Da} S}{2d_ud_v} =\mathrm{Da} \sqrt{\frac{\Delta_0}{d_ud_v}},
\label{eq:qc}
\end{equation}
see Fig.~\ref{fig:threshold}(b) for the Brusselor kinetics introduced in Sec.~\ref{sec:brusselator}.
We can now infer three necessary conditions~\cite{segelDissipativeStructureExplanation1972}. First, $S>0$ and $T_0<0$ require $a_{11}$ and $a_{22}$ to have opposite signs. We then label the species with $a_{11}>0>a_{22}$, calling them $u$ (the activator) and $v$ (the inhibitor). Second, $\Delta_0>0$ requires the cross-couplings to have opposite signs,
\begin{equation}
    \chi:=-a_{12}a_{21}=\Delta_0+a_{11}|a_{22}|>0 .
\label{eq:chi}
\end{equation}
Third, $S>0$ and $T_0<0$ give $d_v/d_u>|a_{22}|/a_{11}>1$: the inhibitor must have the
larger effective diffusivity.

Dividing Eq.~\eqref{eq:inst} by $d_u$ gives an inequality in $r=\sqrt{d_v/d_u}$, $a_{11}r^2-2\sqrt{\Delta_0} r-|a_{22}|>0$, whose positive root yields, with Eq.~\eqref{eq:chi},
\begin{equation}
    \frac{d_v}{d_u}>\rho_c =\left(\frac{\sqrt{\Delta_0}+\sqrt{\chi}}{a_{11}}\right)^{2} .
\label{eq:rhoc}
\end{equation}
The kinetics thus enter through the single number
$\rho_c>1$, and the flow through the ratio $d_v/d_u$,
which depends on $\mathrm{Pe} \ell$ and $\delta$ only
[Eq.~\eqref{eq:dudv}]. See Sec.~\ref{sec:results_shear} for its physical implications.

\subsubsection{Fastest-growing mode}
Beyond the instability onset, a band of wavenumbers is unstable, and the pattern that emerges from small perturbations is dominated by the mode that grows fastest. We therefore look for the maximum of $\mathrm{Re} \lambda_+$ over $s=q^2>0$, for $d_v>d_u$ as required from the pattern criterion. Writing
$\lambda_+=\tfrac{1}{2}[\mathrm{Tr}(\boldsymbol{J}_q) +  R_s]$ with $R_s=\sqrt{\mathrm{Tr}(\boldsymbol{J}_q)^2
-4\det(\boldsymbol{J}_q)}$, Eqs.~\eqref{eq:trdet-gen} and \eqref{eq:chi} give
\begin{equation}
    R_s^2= \left[\mathrm{Da}(a_{11}-a_{22})
    +(d_v-d_u)s \right]^2 - 4\mathrm{Da}^2 \chi.
\label{eq:Rs}
\end{equation}
The term in brackets is positive and increases with $s$, so
$R_s^2$ can change sign at most once, from negative to
positive. This separates two regimes [see Fig.~\ref{fig:threshold}(c) for the Brusselor kinetics introduced in Sec.~\ref{sec:brusselator}]. For $R_s^2<0$, the eigenvalues are complex and $\mathrm{Re} \lambda_+=\mathrm{Tr}(\boldsymbol{J}_q)/2
=(\mathrm{Da} T_0-s \Sigma)/2$ decreases linearly with
$s$, so the maximum cannot be there. For $R_s^2>0$, the
eigenvalues are real, remain real for all larger $s$, and
differentiating $\lambda_+$ twice gives
\begin{equation}
    \frac{\mathrm{d}^2\lambda_+}{\mathrm{d}s^2}
    =-\frac{2\mathrm{Da}^2\chi(d_v-d_u)^2}{R_s^3} < 0 .
\label{eq:concave}
\end{equation}
The growth rate is thus a strictly concave function of
$s$: it has at most one stationary point, and this point is
its maximum. Solving $\mathrm{d}\lambda_+/\mathrm{d}s=0$ therefore yields the fastest-growing (selected) wavenumber and its growth rate,
\begin{subequations}
\begin{align}
    q_{\mathrm{sel}}^2 &= \mathrm{Da}
    \frac{\sqrt{\chi}(d_u+d_v)-(a_{11}-a_{22})\sqrt{d_u d_v}}
     {(d_v-d_u)\sqrt{d_ud_v}}, \label{eq:qsel-gen}\\
    \lambda_{\max} &= \mathrm{Da}
    \frac{a_{11}d_v-a_{22}d_u-2\sqrt{\chi d_ud_v}}{d_v-d_u} .
\label{eq:lmax-gen}
\end{align}
\end{subequations}
Equation~\eqref{eq:qsel-gen} describes a mode only if its
right-hand side is positive, which requires
\begin{equation}
    a_{11}-a_{22}< 2\sigma \sqrt{\chi}, \qquad \sigma=\frac{d_u+d_v}{2\sqrt{d_u d_v}}
    =\frac{1}{2} \left(r+ \frac{1}{r} \right) \ge 1 .
\label{eq:sigma}
\end{equation}
Otherwise the stationary point lies at $s\le0$, the growth
rate decreases for all $s>0$, and no finite-$q$ mode grows
faster than the uniform one.

\subsubsection{Competition between homogeneous and
finite-wavelength modes}
To derive the threshold for the finite-wavelength
instability, we assumed that the homogeneous state remains
stable. For $T_0>0$, the uniform mode is itself unstable,
and a negative determinant no longer tells us that a pattern forms, only that a finite-$q$ mode is also unstable. We therefore ask whether some mode $q>0$ grows faster than the uniform one, i.e., whether $\lambda_{\max}$
[Eq.~\eqref{eq:lmax-gen}] exceeds $\mathrm{Re} \lambda_+(0)$. Since $T_0^2-4\Delta_0=(a_{11}-a_{22})^2-4\chi$, the uniform eigenvalues $\lambda_\pm(0)$ are real for
$a_{11}-a_{22} \ge 2\sqrt{\chi}$ and complex otherwise, and
the two cases have to be treated separately.

In the first case, $a_{11}-a_{22}\ge2\sqrt{\chi}$, the
growth rate $\lambda_+$ is real and concave for all
$s \ge 0$ [Eq.~\eqref{eq:concave}]. The fastest mode is then the selected one if it exists, and the uniform one
otherwise. A finite-$q$ mode thus grows faster than the
uniform one if $q_{\mathrm{sel}}^2>0$, i.e.,
$a_{11}-a_{22}<2\sigma\sqrt{\chi}$ [Eq.~\eqref{eq:sigma}]. Equivalently, the slope of $\lambda_+$ at $s=0$ is positive, $\mathrm{Da} S>\lambda_+(0) \Sigma$.

In the second case, $a_{11}-a_{22}<2\sqrt{\chi}$, the
uniform eigenvalues are complex and $\mathrm{Re}\,\lambda_+(0)=\mathrm{Da}\,T_0/2$. Starting from $s=0$, the growth rate first decreases with slope $-\Sigma/2$, so the slope criterion fails, and yet a finite-$q$ mode can grow faster once the eigenvalues turn real at larger $s$ [see Fig.~\ref{fig:threshold}(c)  for the Brusselor kinetics introduced in Sec.~\ref{sec:brusselator}]. Since $\sigma\ge1$, the selected mode always exists here [Eq.~\eqref{eq:sigma}], and Eq.~\eqref{eq:lmax-gen} gives
\begin{equation}
    \lambda_{\max}-\frac{\mathrm{Da} T_0}{2}
    =\mathrm{Da}\frac{(a_{11}-a_{22})(d_u+d_v)
    -4\sqrt{\chi d_ud _v}}{2(d_v-d_u)},
\label{eq:lmax-diff}
\end{equation}
which is positive for $a_{11}-a_{22} > 2\sqrt{\chi}/ \sigma$.

Both cases combine into a single criterion: a finite-$q$
mode outgrows the uniform one if and only if
\begin{equation}
    \frac{2\sqrt{\chi}}{\sigma} < a_{11}-a_{22}
    < 2 \sigma \sqrt{\chi}.
\label{eq:window}
\end{equation}
The quantities in Eq.~\eqref{eq:window} can be directly interpreted. The kinetics enter through
$a_{11}-a_{22}=a_{11}+|a_{22}|$, the sum of the
self-enhancement rate of the activator and the self-decay rate of the inhibitor, and through $\sqrt{\chi}$, the strength of the feedback loop between the two species. The flow enters only through $\sigma$, which measures the
contrast between the two effective diffusivities,
$\sigma=1$ for $d_u=d_v$, and $\sigma$ grows with
$d_v/d_u$. Like $d_v/d_u$, $\sigma$ depends only on
$\mathrm{Pe} \ell$ and $\delta$ [Eq.~\eqref{eq:dudv}]. At
$\mathrm{Pe} \ell=\sqrt{105\delta/2}$, where shear
equalizes the effective diffusivities (Sec.~\ref{sec:results_shear}), $\sigma=1$ and the window
closes: without a contrast in effective diffusivities, no
finite-$q$ mode can outgrow the uniform one.

\subsection{Role of the Damk{\"o}hler number} 
It is worth mentioning that $\mathrm{Da}$ can be scaled out of the reduced linear problem. Equation~\eqref{eq:Jq-gen} gives
\begin{equation}
    \boldsymbol{J}_q=\mathrm{Da}\left[\boldsymbol{J}-\frac{q^2}{\mathrm{Da}} \begin{pmatrix} d_u & 0 \\ 
    0 & d_v\end{pmatrix} \right] =\mathrm{Da}\;\boldsymbol{J}_{q/\sqrt{\mathrm{Da}}} \Big|_{\mathrm{Da}=1},
\label{eq:Da-J}
\end{equation}
or the Jacobian at $\mathrm{Da}=1$ evaluated at the wavenumber $q/\sqrt{\mathrm{Da}}$. The eigenvalues therefore satisfy
\begin{equation}
    \lambda_\pm(q;\mathrm{Da}) =\mathrm{Da} \lambda_\pm \left(q/\sqrt{\mathrm{Da}};1 \right).
\label{eq:Da-lam}
\end{equation}
Increasing $\mathrm{Da}$ thus stretches the dispersion
relation. All growth rates are multiplied by $\mathrm{Da}$
and all wavenumbers divided by $\sqrt{\mathrm{Da}}$, but the sign of every growth rate, and which mode grows fastest, is unchanged. Accordingly, Eqs.~\eqref{eq:uniform-stab}, \eqref{eq:rhoc} and \eqref{eq:window} do not contain $\mathrm{Da}$, whereas
$q_c, q_{\mathrm{sel}} \propto\sqrt{\mathrm{Da}}$ and
$\omega_H \propto \mathrm{Da}$.

\subsection{The Brusselator model \label{sec:brusselator}}
To apply our theoretical framework, we consider the
Brusselator~\cite{%
prigogineSymmetryBreakingInstabilitiesDissipative1967,%
prigogineSymmetryBreakingInstabilities1968,%
lefeverChemicalInstabilitiesSustained1971,%
auchmutyBifurcationAnalysisNonlinear1975,%
verdascaReentrantHexagonalTuring1992},
a minimal activator ($c_1$)--inhibitor ($c_2$) model with
local self-enhancement and long-range inhibition: 
\begin{align}
\boldsymbol{R}= \begin{pmatrix}
    k_1 a-k_2b c_1 + k_3 c_1^2c_2 - k_4 c_1 \\
    k_2 b c_1-k_3 c_1^2c_2
    \end{pmatrix}.
\label{eq:brusselator_kinetics}
\end{align}
Here, $a$ and $b$ are externally maintained concentrations,
$k_1$ and $k_2$ are the production rates of the two species, and $k_3$ quantifies their non-linear interaction, which produces the activator and consumes the inhibitor. The degradation rate $k_4$ of the activator sets the characteristic time scale of the kinetics,
$\tau_{\rm chem}=1/k_4$. Rescaling as in Sec.~\ref{sec:model} with the base concentration $c_0=\sqrt{k_4/k_3}$, we have
\begin{subequations}
\begin{align}
    \partial_{t'}{c}_1' + W_{x'}'\partial_{x'}c_1'
    &= \frac{1}{{\rm Pe}_1} \partial_{x'}^2c_1'
 + \frac{1}{\epsilon_1}\partial_{y'}^2c_1'
 +{\rm Da} f(c_1',c_2'),
    \label{eq:brusselator_nd_1}\\
    \partial_{t'}{c}_2' + W_{x'}'\partial_{x'}c_2'
    &= \frac{1}{{\rm Pe}_2} \partial_{x'}^2c_2'
     + \frac{1}{\epsilon_2}\partial_{y'}^2c_2'
     + {\rm Da} g(c_1',c_2'),
\label{eq:brusselator_nd_2}
\end{align}
\end{subequations}
with $A=k_1a/(k_4c_0)$ and $B=k_2b/k_4$, which measure how
fast both species are produced relative to the degradation
rate of the activator. We further used $\delta_{1i}/\mathrm{Pe}_1=1/\mathrm{Pe}_i$
and $\delta_{1i}/\epsilon_1=1/\epsilon_i$ to write the
diagonal entries of $\boldsymbol{D}'$ species by species.  The reaction vector $\boldsymbol{R}'=(f,g)^\top$ becomes 
\begin{subequations}\label{eq:fg-bru}
    \begin{align}
        f(c_1',c_2') &= A -(B+1)c_1' + (c_1')^2 c_2', \\
        g(c_1',c_2') &= Bc_1' - (c_1')^2 c_2'.
    \end{align}
\end{subequations}
In the reduced description [Eqs.~\eqref{eq:red}], the equations for the cross-sectionally averaged concentrations, $u=\bar{c}_1'$ and $v=\bar{c}_2'$, read: 
\begin{subequations}
\begin{align}
  \partial_t u &= d_u \partial_\xi^2 u + {\rm Da}\big[A-(B+1) u+u^2v\big], \label{eq:redu}\\
  \partial_t v &= d_v \partial_\xi^2 v + {\rm Da}\big[B u-u^2v\big].       \label{eq:redv}
\end{align}
\end{subequations}
The stationary state of the Brusselator kinetics is $(u^\star,v^\star)=(A,B/A)$ and we have the quantities:
\begin{equation}
    \boldsymbol{J}=\begin{pmatrix} B-1 & A^2\\ -B & -A^2\end{pmatrix}, \qquad
\begin{aligned}
    T_0 &=B-1-A^2, &\Delta_0 &=A^2, \\
    \chi &=A^2B,  &a_{11}-a_{22} &=A^2+B-1.
    \end{aligned}
\label{eq:J-bru}
\end{equation}
Equations~\eqref{eq:Jq-gen}--\eqref{eq:S-gen} become
\begin{equation}
    \boldsymbol{J}_q=\begin{pmatrix} \mathrm{Da}(B-1)-q^2d_u & \mathrm{Da} A^2 \\ 
    -\mathrm{Da}\,B & -\mathrm{Da}\,A^2-q^2d_v
\end{pmatrix},
\label{eq:Jq-bru}
\end{equation}
and
\begin{subequations}\label{eq:trdet-bru}
\begin{align}
\mathrm{Tr}(\boldsymbol{J}_q)&=\mathrm{Da}\,(B-1-A^2)-q^2(d_u+d_v),\\
\det(\boldsymbol{J}_q)&=\mathrm{Da}^2A^2-\mathrm{Da}\,S\,q^2
+q^4d_ud_v,
\end{align}
\end{subequations}
with $S=(B-1)d_v-A^2d_u$. We now evaluate the general
conditions for the system's stability. The resulting phases are summarized in Table~\ref{table}.

\subsubsection{Hopf threshold}
Since $\Delta_0=A^2>0$, the Hopf bifurcation $T_0=0$ occurs at
\begin{equation}
    B_H=1+A^2,
\label{eq:b_hopf}
\end{equation}
with frequency $\omega_H=\mathrm{Da} A$, independently of
the flow.

\subsubsection{Finite-wavelength instability}
At the onset, Eq.~\eqref{eq:inst} reads
$(B-1)d_v-A^2d_u=2A\sqrt{d_ud_v}$, and solving for $B$ gives the effective Turing threshold above which a finite-wavelength instability emerges,
\begin{equation}
    B_T^{\mathrm{eff}} =\left(1 + A\sqrt{\frac{d_u}{d_v}} \right)^{2},
\label{eq:turing_threshold_effective}
\end{equation}
with the onset wavenumber $q_c^2=\mathrm{Da} A /\sqrt{d_u d_v}$ [Eq.~\eqref{eq:qc}]. Equivalently, Eq.~\eqref{eq:rhoc} reads $d_v/d_u>\rho_c=A^2/(\sqrt B-1)^2$. Here, the threshold $B_T^{\mathrm{eff}}$ depends on the flow via $d_u$ and $d_v$.

\subsubsection{Competition between homogeneous and
finite-wavelength modes}
Above the Hopf threshold, $B>B_H$, the uniform mode is
unstable, and we use Eq.~\eqref{eq:window} to decide
whether a finite-$q$ mode still grows faster. With
Eq.~\eqref{eq:J-bru}, this criterion reads
\begin{equation}
    \frac{2A \sqrt B}{\sigma}< A^2+B-1 < 2\sigma A\sqrt B .
\label{eq:window-bru}
\end{equation}
The eigenvalues $\lambda_\pm(0)$ are complex for $A^2+B-1<2A\sqrt B$, i.e. in the window
$(A-1)^2<B<(A+1)^2$, which contains $B_H$. There, only the
lower bound of Eq.~\eqref{eq:window-bru} matters. At
$B=B_H$ it reads $A\sqrt{\sigma^2-1}>1$, i.e., $A>A_c$,
with $A_c$ the Turing--Hopf point [Eq.~\eqref{eq:turing-hopf_point}], and it remains satisfied for all larger $B$. For $A>A_c$, a finite-$q$ mode therefore grows the fastest throughout this window. Above it, $B>(A+1)^2$, only the upper bound matters. The equality, $A^2+B-1=2\sigma A\sqrt B$, is a quadratic equation in $\sqrt{B}$ whose upper root is the crossover
\begin{equation}
    \sqrt{B_\star}=\frac{A}{2} \left( r+\frac{1}{r} \right)
    + \sqrt{1+\frac{A^2}{4} \left(r-\frac{1}{r}\right)^{2}}, \qquad r= \sqrt{d_v/d_u}.
\label{eq:Bstar}
\end{equation}
For $B_H < B <B_\star$ both modes are unstable and a
finite-$q$ mode grows the fastest, for $B>B_\star$ the uniform mode grows the fastest. For large $A$, Eq.~\eqref{eq:Bstar} reduces to $B_\star\simeq A^2d_v/d_u$.

\begin{table}[htbp]
\begin{center}
\begin{tabular}{@{}l|l|l|l@{}}
\toprule
$B$ & unstable modes & fastest mode & linear prediction \\
\midrule
$<B_T^\mathrm{eff}$  & ---        & ---        & homogeneous \\
($B_T^\mathrm{eff}$, $B_H$)  & finite-$q$ & finite-$q$ & \textbf{pattern} \\
($B_H$, $(A+1)^2$)  & finite-$q$ and $q=0$ & finite-$q$\,$^\dagger$ & \textbf{competition}\\
($(A+1)^2$, $B_\star$)  & finite-$q$ and $q=0$ & finite-$q$ & \textbf{competition}\\
$>B_\star$                & finite-$q$ and $q=0$           & $q=0$    & uniform oscillation \\
\bottomrule
\end{tabular}
\end{center}
\caption{Summary of the different phases depending on the
reaction parameter $B$ predicted by the one-dimensional
theory, for $A>A_c$ (so that $B_T^\mathrm{eff}<B_H$).
$^\dagger$In this window the uniform eigenvalues are
complex and the slope criterion does not apply. The fastest
mode follows from the lower bound of Eq.~\eqref{eq:window-bru}. \label{table}}
\end{table}

\subsection{Numerical linear theory for the 2D system}
\label{sec:numerical_linear}
To get more insight into the physics of the system and further validate our model, we can alternatively linearize the full two-dimensional system directly, without invoking the one-dimensional reduction, by using the single axial mode $\delta \boldsymbol{w} = \big(\delta u(y), \delta v (y) \big)^\top e^{\lambda t + i q x}$, where $y \in [-1,1]$ is the rescaled transverse coordinate (we ignore the boundary conditions at the in- and outlet here). We note that here, $\delta u$ and $\delta v$ denote the full two-dimensional fields. In the lab frame, with $W_x(y) = \frac{3}{2}(1-y^2)$, we have:
\begin{subequations}
\begin{align}
  \lambda \delta u &= -iq W_x(y) \delta u
     + \frac{1}{\mathrm{Pe}}\Bigl(-q^2 \delta u + \frac{\partial_y^2 \delta u}{\ell^2}\Bigr)
     + {\rm Da}\bigl[(B-1)\delta u + A^2 \delta v\bigr], \label{eq:exactu}\\
  \lambda \delta v &= -iq W_x(y) \delta v
     + \frac{\delta}{\mathrm{Pe}}\Bigl(-q^2 \delta v + \frac{ \partial_y^2 \delta v}{\ell^2}\Bigr)
     + {\rm Da}\bigl[-B \delta u - A^2 \delta v\bigr].\label{eq:exactv}
\end{align}
\end{subequations}
Discretizing $y$ on $N_y$ points (with $N_y=64$ for our results) turns this into a dense $2N_y\times2N_y$ complex eigenvalue problem, which we solve numerically by constructing the matrix $\mathcal{L}$ such that $\lambda (\delta u, \delta v) = \mathcal{L} (\delta u, \delta v)$.  We note that besides the linearization, Equations~\eqref{eq:exactu}--\eqref{eq:exactv} make no approximation, whereas in our effective one-dimensional theory the eigenvalue $\lambda_\pm(q)$ is the leading term of an expansion in $\epsilon=\mathrm{Pe}\ell^2$. 

In the two-dimensional stability analysis, each point of the $(A,B)$ plane is then classified, as for the reduced theory, by comparing the growth rate of the uniform mode, $\mathrm{Re} [\lambda(0)]$, with that of the fastest finite-wavenumber mode, $\max_{q>0}\mathrm{Re}[\lambda(q)]$: the system is stable if both are negative, the Turing phase emerges if only the latter is positive, a Hopf phase exists if the uniform mode is unstable and grows fastest, and the competition phase results if both are unstable and a finite-wavenumber mode grows faster.  This is in line with the strategy originally used by Vasquez~\cite{vasquezChemicalInstabilityInduced2004} in his numerical stability analysis. This approach here serves as an independent benchmark to validate the closed-form thresholds of our effective one-dimensional theory.

Lastly, we note that we also solve the full non-linear equations [Eqs.~\eqref{eq:brusselator_nd_1}--\eqref{eq:brusselator_nd_2}] numerically, starting from the homogeneous state perturbed by small random noise. The numerical method and how phases and wavelengths are read off the simulations are described in Appendix~\ref{app:numerics}.

\section{Results \label{sec:results}}
Here, we first discuss our predictions for the impact of flow on pattern formation in confined channel geometries (Sec.~\ref{sec:results_shear}). In particular,  we provide the full phase diagram predicted by our reduced theory and compare it to both the numerical two-dimensional stability analysis and the non-linear numerical results (Sec.~\ref{sec:results_phases}). Our non-linear simulations provide insights into the emergent oscillating Hopf phases and Turing patterns. Finally, we quantify the wavelength of the latter using our reduced theory (Sec.~\ref{sec:results_wavelength}). 

\subsection{Shear-induced Turing patterns in long slender channels \label{sec:results_shear}}
\begin{figure}[htp]
    \centering
    \includegraphics[width=0.8\columnwidth]{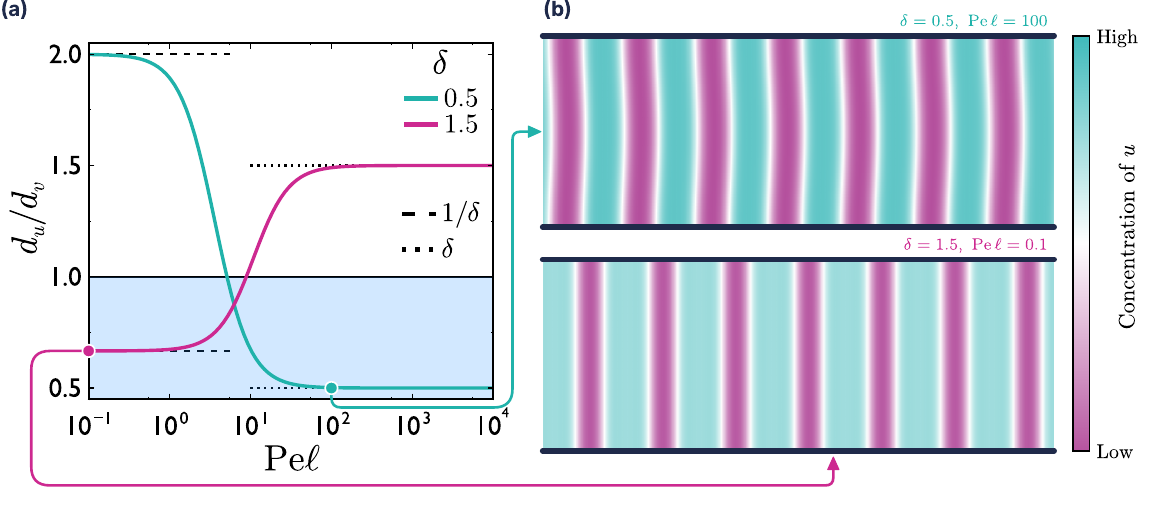}
    \caption{Shear-induced patterns. (a) Ratio of the effective (Taylor--Aris) diffusivities $d_u/d_v$ as a function of $\mathrm{Pe} \ell$ for $\delta=D_v/D_u=0.5$ and $1.5$. The ratio approaches $1/\delta$ at small $\mathrm{Pe} \ell$ (dashed), corresponding to the classical Turing instability in the absence of flow. It converges to $\delta$ at large $\mathrm{Pe} \ell$ (dotted) and depends on $\mathrm{Pe}$ and $\ell$ only through $\mathrm{Pe} \ell$. The shaded region $d_u/d_v<1$ is where a Turing pattern can exist.
    (b)~Simulated activator field at the two marked points ($A=7$, $B=49$). Low concentrations are indicated in purple and high concentrations in teal. Top: $\delta=0.5$ at $\mathrm{Pe} \ell=100$ ($\mathrm{Pe}=10^6$, $\ell=10^{-4}$). The flow has reversed the diffusivity ordering and creates stripes that are bent by the Poiseuille profile and drift with the mean flow. Bottom: $\delta=1.5$ at $\mathrm{Pe} \ell=0.1$ ($\mathrm{Pe}=10^2$, $\ell=10^{-3}$). Here, shear is negligible and the classical Turing pattern is recovered, with straight stripes.}
    \label{fig:ratio_dudv}
\end{figure}

In our reduced theory, the stability analysis has revealed that advection alone does not shift the thresholds. The impact of the Poiseuille flow enters due to its associated shear ($\propto \partial_y W_x$) in the threshold for a finite-wavelength instability~$B_T^{\rm eff}$, corresponding to an effective Turing threshold. Most importantly, to leading order in $\epsilon=\mathrm{Pe}\ell^2$, the molecular diffusivities of the chemical species become replaced by effective diffusivities, which are well-known from classical Taylor--Aris dispersion.  For strong flows, the latter grow as $D_{\mathrm{eff},i} \propto H^2 W_0^2/D_i$, showing that flows can enhance the diffusivities of molecular species and are more important for slower-diffusing species. This feature enters the Turing threshold [Eq.~\eqref{eq:turing_threshold_effective}] as: 
\begin{align}
    B_T^\mathrm{eff} &= \left( 1+ A \sqrt{\frac{d_u}{d_v}} \right)^2= \left( 1 + A \sqrt{\frac{\delta \left( 105 + 2\ell^2\,\mathrm{Pe}^2 \right)}{105\,\delta^2 + 2\ell^2\,\mathrm{Pe}^2}} \right)^{2}.\label{eq:turing_threshold_effective_Pe}
\end{align}
Thus, patterns can exist within the window $B_T^{\rm eff}<B<B_H=1+A^2$, which is only possible if the ratio of effective diffusivities is small $d_u/d_v<1$. For weak flows $ \mathrm{Pe} \ell \ll 1$, the threshold approaches the classical Turing threshold: $B_T \approx ( 1+ A /\sqrt{\delta} )^2$ with $\delta=D_v/D_u$~\cite{murrayMathematicalBiology1989}. This means that patterns are only possible if $1/\sqrt{\delta}<1$, reflecting that the inhibitor has to diffuse much faster than the activator: $D_v>D_u$. Conversely, for strong flows $ \mathrm{Pe} \ell \gg 1$, we find $B_T \approx ( 1+A \sqrt{\delta})^2$, showing that strong flows flip the criterion for patterns, $\sqrt{\delta}<1$ or $D_u>D_v$. This suggests that in systems where the activator diffuses faster than the inhibitor, patterns are possible in the presence of strong flows. 

Figure~\ref{fig:ratio_dudv} shows the ratio of effective diffusivities $d_u/d_v$ for different ratios of molecular diffusivities $\delta=0.5$ and $\delta=1.5$ as a function of flow strength Pe$\ell$. The two effective diffusivities are equal ($d_u=d_v$) at $(\mathrm{Pe}\ell)^\ast=\sqrt{105 \delta/2}$. At this crossover the shear
reverses the ordering of the effective diffusivities. For $\delta=1$ there is no crossover since both species are dispersed equally, $d_u/d_v = 1$ for every $\mathrm{Pe}\ell$ [Eq.~\eqref{eq:turing_threshold_effective_Pe}], and the flow cannot create a pattern. The region of the parameter space where a Turing pattern can exist is indicated in light blue. For $\delta=1.5$ our reduced theory predicts that a  stable pattern at weak flows Pe$\ell \lesssim 10$, disappears at strong flows, suggesting that flows can destroy Turing patterns that are stable in their absence. Conversely, we find that for $\delta=0.5$ stable patterns emerge as Pe$\ell\gtrsim 10$. We validate these predictions by full non-linear simulations of the two-dimensional system. Interestingly, for weak flows $\mathrm{Pe} \ell \gtrsim 0.1$, and $\delta=1.5$ the Brusselator kinetics leads to stripe patterns in the channel geometry. Similar patterns are observed for $\delta=0.5$ at much stronger flows, $ \mathrm{Pe} \ell\gtrsim 10$. The latter stripes show weak bending resulting from the applied Poiseuille flow profile.

We note that both patterns are (almost) uniform across the channel (in $y$), suggesting that our assumptions for the reduced theory are fulfilled. 

\subsection{Phase diagrams in the presence and absence of flows \label{sec:results_phases}}
\begin{figure}[htp]
    \centering
    \includegraphics[width=0.75\columnwidth]{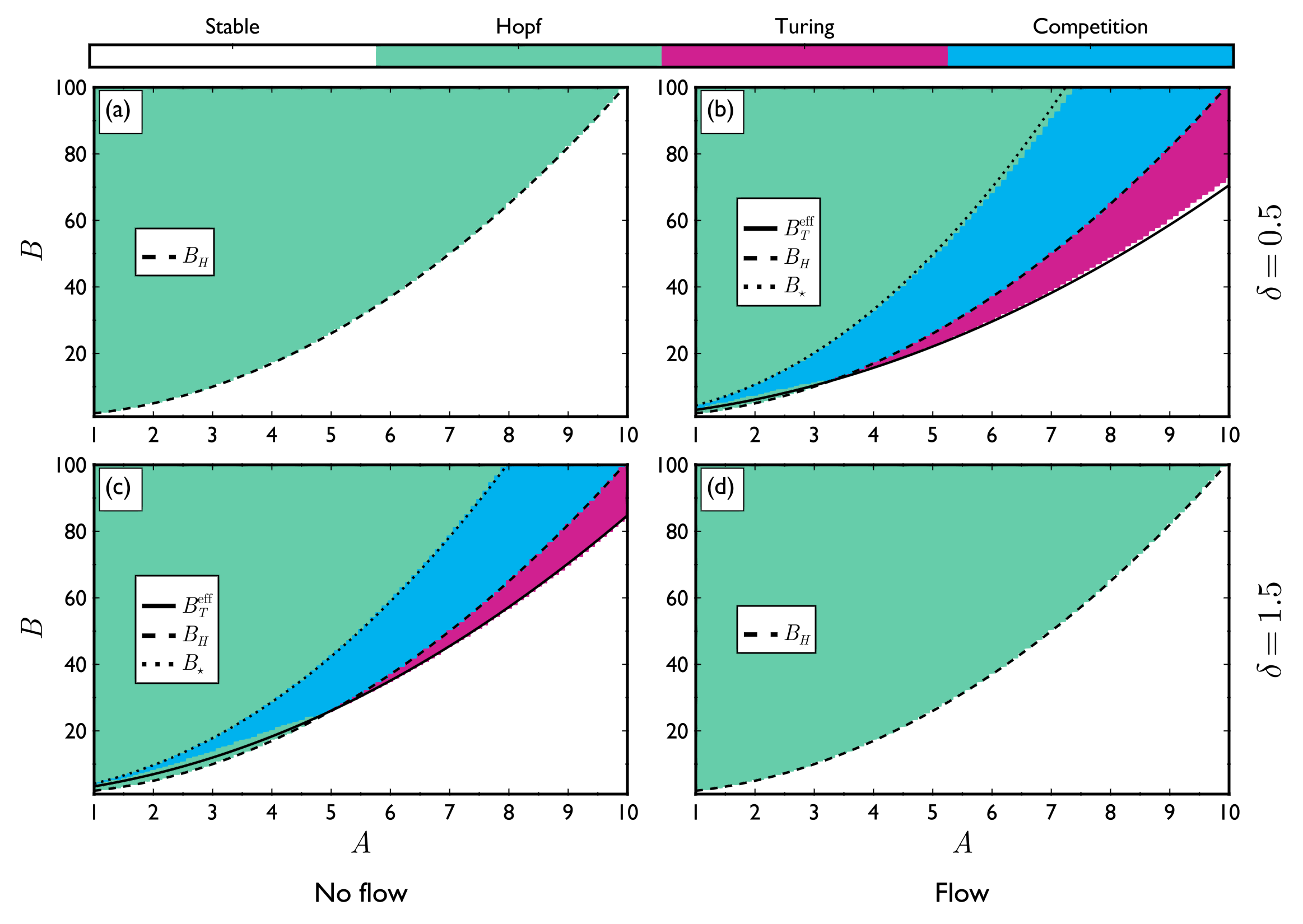}
    \caption{Phase diagrams in the $(A,B)$ plane (left) without and (right) with flow ($\mathrm{Pe} \ell=20$), for $\delta=0.5$ (top) and $\delta=1.5$ (bottom). Here, $\mathrm{Da}=1$ and $\epsilon=0.01$. Lines correspond to predictions from the linear theory of the reduced equations, classified by the fastest-growing mode. Lines: Hopf threshold $B_H$ (dashed) [Eq.\eqref{eq:b_hopf}], effective Turing threshold $B_T^\mathrm{eff}$ (solid) [Eq.~\eqref{eq:turing_threshold_effective}], crossover $B_\star$ (dotted) [Eq.~\eqref{eq:Bstar}]. Colors indicate the results from the numerical linear stability analysis of the two-dimensional system.}
    \label{fig:phase_nf_f}
\end{figure}
To quantify the different phases and the impact of flows, we present the phase portrait in the $(A,B)$ space and compare the flow and no-flow situations in Fig.~\ref{fig:phase_nf_f}.
At $\delta=0.5$ (first row), the classical Brusselator can only be in a stable or oscillatory (Hopf) state and does not exhibit any patterns. In contrast, turning on the flow (right column) with $\mathrm{Pe}\ell \geq 10$ makes a Turing band appear and patterns can form. Since the finite-$q$ mode is also unstable, we additionally note a ``competition'' zone where Hopf and Turing states decide the system's state. When the inhibitor is more diffusive than the activator [Fig.~\ref{fig:phase_nf_f}, second row] ($\delta =1.5$), the Brusselator can display patterns at rest. In the presence of flow at $\mathrm{Pe}\ell =20$, shear reverses the effective diffusivity ordering, and the Turing band and the competition region disappear, leaving only stable and oscillating states.

\begin{figure}[tp]
    \centering
    \includegraphics[width=0.75\linewidth]{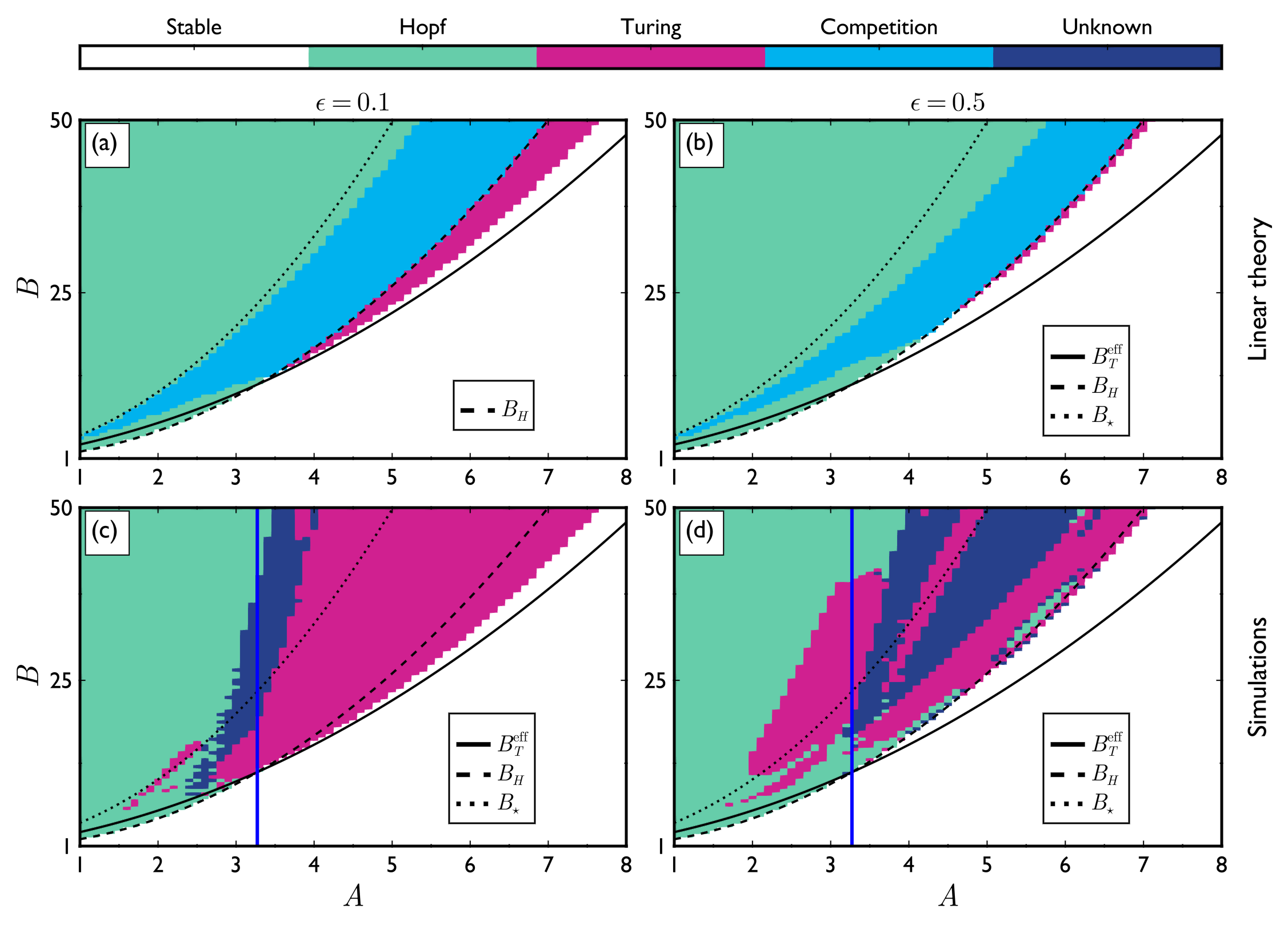}
    \caption{Validation of the reduced theory at $\mathrm{Pe} \ell=20$, $\delta=0.5$, and $\mathrm{Da}=1$ for $\epsilon=\mathrm{Pe} \ell^2=0.1$ (left) and $\epsilon =0.5$ (right). (a)--(b): phases from the exact linear operator of the two-dimensional channel, classified by the fastest-growing mode. (c)--(d): phases read off noise-seeded simulations from the saturated state. The vertical line is the Turing--Hopf point $A_c=3.27$, Eq.~\eqref{eq:turing-hopf_point}.}
    \label{fig:comparison_pp}
\end{figure}

We assess the validity of the predictions of our reduced model by comparing to the phases obtained by the numerical two-dimensional stability analysis (shown in color in Fig.~\ref{fig:phase_nf_f}). For this set of parameters both predictions are very close as the perturbation parameter is small ($\epsilon =\mathrm{Pe}\ell^2= 0.01$), confirming that for very narrow channels our analytical predictions are accurate. 

To study systematically how well the reduced theory organizes the instability across the parameter space, we construct the phase portraits for different $\epsilon=0.1, 0.5$ in Fig.~\ref{fig:comparison_pp}. At $\epsilon =0.1$, all three boundaries closely track the transitions found in the exact linear operator. The same region ordering is recovered in the direct simulations (panel c) but with one difference: the competition region does not appear as a distinct phase there since the simulation determines which state is realized by evolving the system from a perturbed homogeneous state to its final state, which is either homogeneous, patterned, or oscillating. To compare to the theory, we determine the Turing--Hopf point from $B_H = B_T^\mathrm{eff}$. For $d_u/d_v<1$ it evaluates to: 
\begin{equation}
    A_c  = 2 \dfrac{\sqrt{d_vd_u}}{d_v-d_u}. \label{eq:turing-hopf_point}
\end{equation}
The Turing band ($B_T^\mathrm{eff} < B < B_H$) can only exist for $A> A_c$. The transition between an oscillating and a patterned state resulting from the non-linear numerics can be accurately predicted by $A_c$, which is indicated as a vertical blue line in Fig.~\ref{fig:comparison_pp}. 

Besides stable, Hopf and Turing states, the simulations show a fourth state that fits neither class: a uniform oscillation that coexists with a stationary, localized Turing domain, so that the spatial and temporal spreads of $\langle u\rangle$ are comparable. These states lie near $A_c$, where the uniform and the finite-wavenumber modes are both unstable with comparable growth rates. Their existence cannot be decided by linear theory, which only orders growth rates, but are expected from a weakly non-linear analysis point of view (see Refs.~\cite{dewitSpatiotemporalDynamicsCodimensiontwo1996,meixnerGenericSpatiotemporalDynamics1997}).

As the transverse diffusion time becomes comparable to the axial solute advection time ($\epsilon = 0.5$), the effective theory can no longer properly predict the threshold above which the Turing band opens: the separation of timescales breaks down. Since the reduction requires $\epsilon \mathrm{Da} \ll 1$, the same breakdown would occur at smaller $\epsilon$ for larger $\mathrm{Da}$. The uniform mode ($q=0$) involves no diffusion and is therefore reproduced exactly by the reduction at any $\epsilon$. The finite-$q$ growth rate, and hence $B_T^\mathrm{eff}$, is only approximated by transverse coarse-graining, and the approximation breaks down as $\epsilon$ grows, because the reduction discards the transverse structure of the perturbation. This same error propagates directly into the Turing--Hopf point: $A_c$ is fixed by the ratio $d_v/d_u$, and depends only on $\mathrm{Pe}\ell$ and $\delta$, not on $\epsilon$. Thus, the Turing phases found in the non-linear simulations cross over the threshold $A_c$ and cannot be described by the reduced model.

\subsection{Analysis of the Hopf and Turing phases \label{sec:results_wavelength}}

\begin{figure*}[htp]
  \centering
  \includegraphics[width=\textwidth]{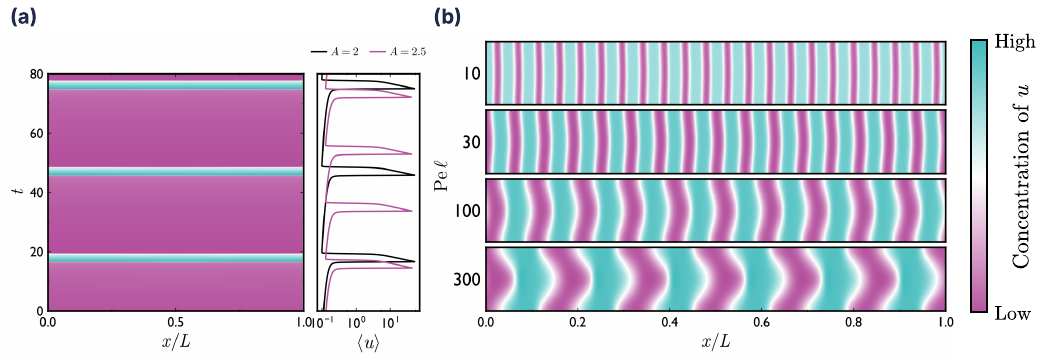}
  \caption{(a) Oscillations of the concentration shown as a kymograph of the cross-sectionally averaged concentration $\langle u \rangle (x,t)$ (left), and as average along the channel length $\langle u \rangle$ against time (right). Parameters are $A=2$, $B=20$, $\delta=0.5$, $\mathrm{Da}=1$, and $\mathrm{Pe}\ell=20$. (b) Turing patterns for different flow strengths $\mathrm{Pe}\ell$. Parameters are $A=7$, $B=49$, $\delta=0.5$, $\mathrm{Da}=1$, and $\ell=0.001$.}
  \label{fig:stripes_hopf}
\end{figure*}

The non-linear simulations allow us to inspect the concentration profiles across the channel. In the Hopf phase, the concentration is uniform across and along the channel and oscillates in time between low and high values, see Fig.~\ref{fig:stripes_hopf}(a). This oscillation does not depend on the flow, and its period scales as $1/\mathrm{Da}$. Close to the onset, the oscillation is small and harmonic, with the predicted frequency $\omega_H=\mathrm{Da}A$ of the linearized kinetics. In Fig.~\ref{fig:stripes_hopf}(a), however, $B=20$ lies far above the onset ($B_H=5$ and $7.25$ for $A=2$ and $2.5$), and the state is a relaxation oscillation whose period is set by the non-linear kinetics: a slow build-up of $v$ (low $u$ concentration), then a fast spike in $u$ (low $v$ concentration). A larger $A$ still produces more spikes, but oscillations are slower than the linear Hopf frequency $\omega_H$, which only describes small oscillations near $B_H$. We note that changing $\mathrm{Da}$ only rescales time, whereas changing $A$ changes the kinetics themselves.

\begin{figure}[tp]
    \centering
    \includegraphics[width=0.8\linewidth]{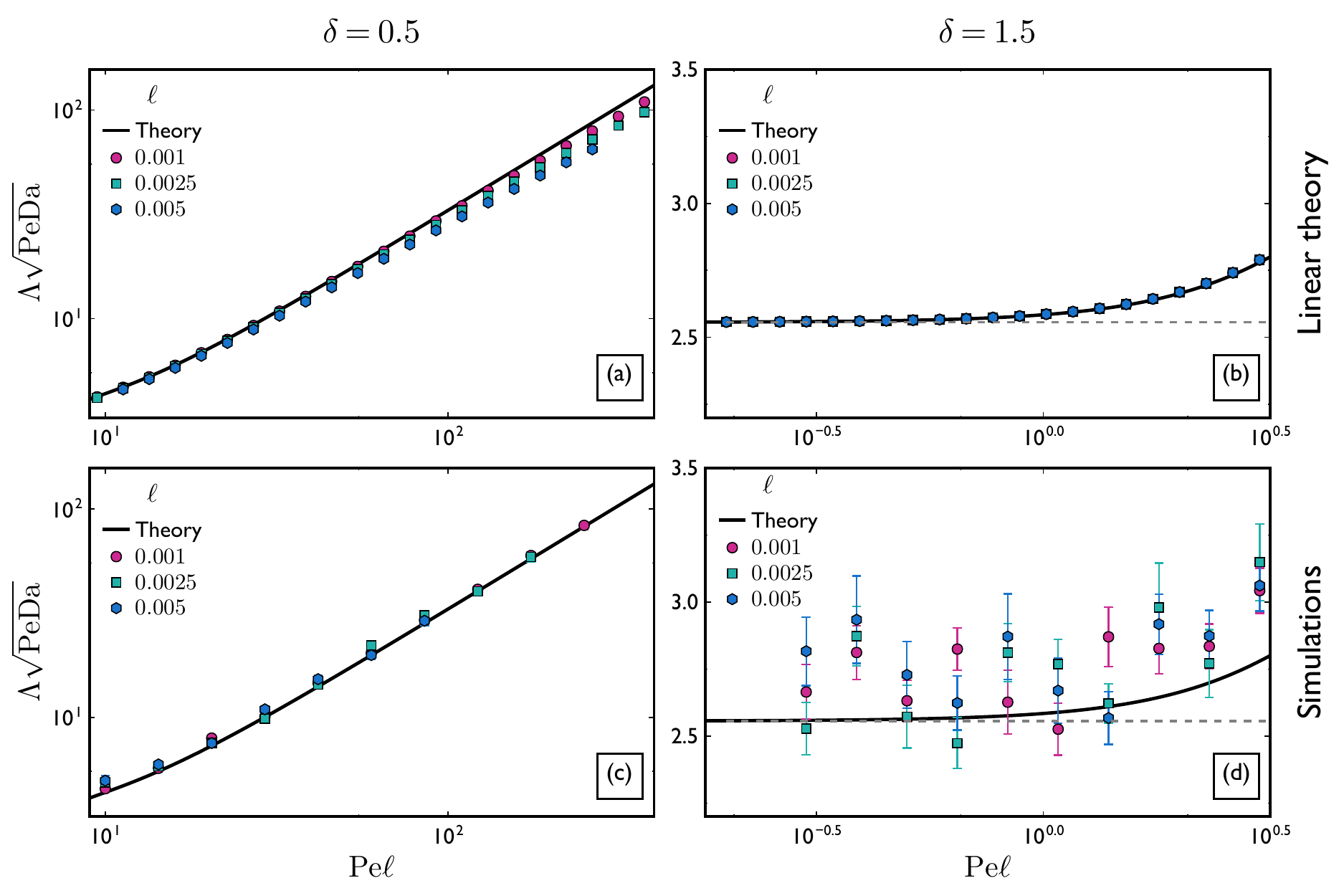}
    \caption{Rescaled wavelength against $\mathrm{Pe} \ell$ for $\delta=0.5$ (left, the flow creates the pattern) and $\delta=1.5$ (right, the flow destroys it), $A=7$, $B=49$, $\mathrm{Da}=1$, for three aspect ratios $\ell$. Solid line: reduced theory [Eq.~\eqref{eq:qsel}], a single curve for all $\ell$. (a)--(b): fastest-growing wavelength of the exact linear operator. (c)--(d): wavelength of the saturated pattern in simulations, shown for $\epsilon \lesssim 0.5$. Only wavelengths $\Lambda=L_x/n$ fit in the periodic box of length $L_x$, with $n$ the number of periods. Bars show $\pm \Lambda/(2n)$, half the spacing to the neighboring admissible wavelengths $L_x/(n \pm1)$. In (b) and (d), the dashed gray line indicates the low P{\'e}clet limit of Eq.~\eqref{eq:low_pe_lim}.}
    \label{fig:wavelength}
\end{figure}

In the Turing phase, our results show stripes that increase in wavelength for increasing flow strength ${\rm Pe\ell}$, see Fig.~\ref{fig:stripes_hopf}(b). These stripes move with the average flow velocity and bend in response to the distortion by the flow profile. The latter is more visible by increasing the flow strength. As summarized in Fig.~\ref{fig:ratio_dudv}, in the presence of flow, patterns can emerge for both $\delta <1$ (at large ${\rm Pe\ell}$) and $\delta >1$ (at small ${\rm Pe\ell}$). We can further quantify these patterns by computing their wavelength, which is obtained from the selected (i.e., the fastest-growing) wavenumber away from the threshold.  For the Brusselator, the selected wavenumber [Eq.~\eqref{eq:qsel-gen}] reads
\begin{equation}
    q_\mathrm{sel}^2 = \frac{{\rm Da}\left[A\sqrt{B}(d_u+d_v) - (A^2+B-1)\sqrt{d_ud_v}\right]}{(d_v-d_u)\sqrt{d_ud_v}}, \label{eq:qsel}
\end{equation}
with associated maximum growth rate [Eq.~\eqref{eq:lmax-gen}]
\begin{equation}
    \lambda_\mathrm{max} = \mathrm{Re}\,\lambda_+(q_\mathrm{sel}) = {\rm Da}\,\frac{\left( A \sqrt{d_u} - \sqrt{Bd_v} \right)^2 - d_v}{d_v - d_u}. \label{eq:lambda_max}
\end{equation}
We note that the selected wavelength can then be written as: 
\begin{align}
\Lambda = \frac{2\pi}{q_{\rm sel}} = \frac{\varphi(\mathrm{Pe} \ell, A, B, \delta) }{\sqrt{\mathrm{Pe}{\rm Da}}},
\end{align}
with some function $\varphi$. Importantly, we observe that there is no explicit dependence on $\ell$, which we can directly test with the numerical stability analysis  of the two-dimensional system and the non-linear numerics.

Figure~\ref{fig:wavelength} shows the rescaled wavelength $ \Lambda\sqrt{\mathrm{Pe} \mathrm{Da}}$ as a function of $\mathrm{Pe}\ell$, for the regime of high P{\'e}clet number and $\delta<1$ and the regime of low P{\'e}clet number and $\delta >1$ (as identified in Fig.~\ref{fig:ratio_dudv}).
The top row compares the theoretical curve against the exact linear operator, which, unlike the reduced theory, depends on $\mathrm{Pe}$ and $\ell$ independently through $1/\mathrm{Pe}$ and $\epsilon= \mathrm{Pe}\ell^2$. The selected wavenumber is read from the maximum of the leading eigenvalue of the operator built from Eqs.~\eqref{eq:exactu}--\eqref{eq:exactv}. In the high--$\mathrm{Pe \ell}$ panel, we find that the data collapse to a single master curve, as suggested by our reduced theory (solid line in Fig.~\ref{fig:wavelength}). The collapse holds over two decades in $\mathrm{Pe}\ell$ and starts to deteriorate at the largest $\mathrm{Pe}\ell$, where the perturbation parameter $\epsilon={\rm Pe}\ell^2$ becomes the largest. This result is a genuine test of the reduction, since $\epsilon$ ranges over more than an order of magnitude across it. By contrast, the low--$\mathrm{Pe}\ell$ panel shows a flat profile of $\Lambda\sqrt{\mathrm{Pe} \mathrm{Da}}$. In the low P{\'e}clet limit, where $d_u/d_v \approx 1/\delta$, we find
\begin{equation}
\Lambda \approx \frac{2\pi}{\sqrt{\mathrm{Pe}{\rm Da}}}\sqrt{ \frac{(\delta-1)\sqrt{\delta}}{A\sqrt{B}(1+\delta) - (A^2+B-1)\sqrt{\delta}}}, \label{eq:low_pe_lim}
\end{equation}
which results in a plateau that is determined by the molecular diffusivity ratio $\delta$, not by the flow. We note that in the plotted range of $\mathrm{Pe}\ell$, $\epsilon$ never exceeds $\approx 0.02$. The agreement here mainly confirms the asymptotic behavior obtained from the reduced model. 

The bottom row compares the same reduced-theory curve against direct simulations where the wavelength is read off as a box mode $q=2\pi n/ L_x$ on a periodic domain of length $L_x$, with resolution $\Delta \Lambda/\Lambda = 1/(2n)$. Since $\Lambda$ grows with $\mathrm{Pe}\ell$ on the $\delta <1$ branch, $n$ would shrink in a box of fixed length and the resolution would degrade. Therefore, we choose $L_x$ between $L$ and $4L$ so that it holds for approximately ten periods, bounding the bars by $5 \%$ of the value. We observe that for high ${\rm Pe}\ell$ (and $\delta <1$) the numerics collapse for all aspect ratios onto a single master curve and nicely agree with the analytical predictions. For the small ${\rm Pe}\ell$ (and $\delta >1$) scenario, the simulated wavelength is systematically longer than the fastest-growing one. In fact, the simulations are offset by the same absolute amount in both cases, but this effect is simply invisible on the logarithmic scale of Fig.~\ref{fig:wavelength}(c). The reduced theory therefore captures qualitatively how the wavelength depends on the flow and on the geometry.

\section{Summary and Conclusions}
\begin{figure}[htp]
    \centering
    \includegraphics[width=\columnwidth]{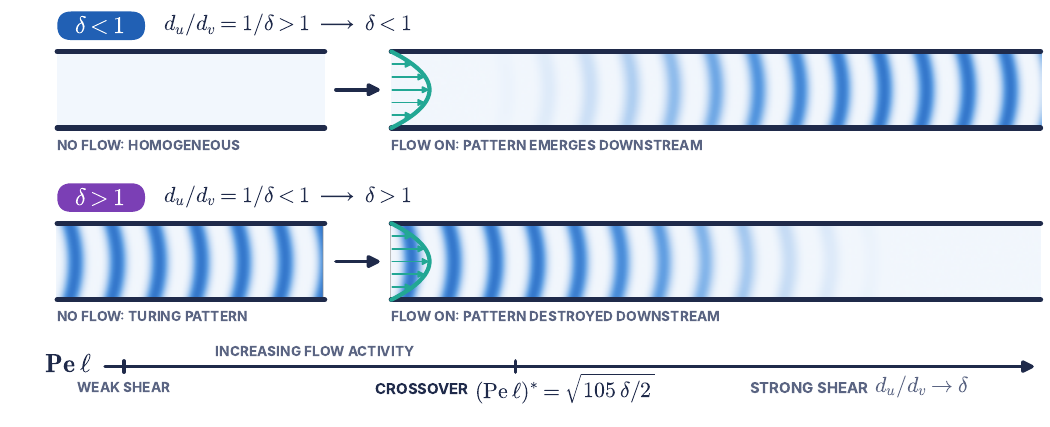}
    \caption{Shear-induced creation and destruction of Turing patterns in a slender channel. Taylor--Aris dispersion enhances the axial diffusivity of the slower species more, so the instability criterion $d_u/d_v$ moves from $1/\delta $ at weak flows to $\delta$ at strong flows.}
    \label{fig:schema1}
\end{figure}

In this work we have shown that a Poiseuille flow in slender channels can act as an independent pattern-forming mechanism in a reaction--diffusion system, capable of both creating and destroying chemical patterns without any change to the underlying reaction kinetics. Studying a Brusselator system advected by a planar flow in a slender channel, we used an invariant-manifold reduction to obtain effective one-dimensional equations for the cross-sectionally averaged concentrations, where the flow, in addition to an average advection,  manifests itself through species-dependent Taylor--Aris diffusivities, $d_u$ and $d_v$. Their ratio can cross one even when the  ratio of their molecular diffusivities $\delta$ violates the classical Turing order, so that shear can open a pattern-forming window for $\delta<1$, where none would exist at rest, and close one for $\delta>1$, where a pattern would otherwise be stable. In the non-linear regime, the pattern forms bands spanning the channel width: the channel is much narrower than the wavelength, so the concentration is nearly uniform across it. When the flow creates the pattern, these bands drift downstream with the mean flow and are bent by the Poiseuille profile. Interestingly, the wavelength of the patterns increases with the P{\'e}clet number, which quantitatively matches the predictions of our reduced theory.

While here we have considered a pressure-driven Poiseuille flow, other profiles, such as recirculating or pulsatile flows, could be relevant for cortex-driven streaming in \textit{Drosophila} or \textit{C. elegans} embryos, as well as streaming in pollen tubes and root hairs~\cite{htetCortexdrivenCytoplasmicFlows2023, htetAnalyticalMethodsCytoplasmic2025}. In addition, the invariant-manifold reduction method has been applied to time-dependent flows~\cite{marbachActiveControlDispersion2019}, which could impact pattern formation in slime molds~\cite{schickDecisionMakingLightTrappedSlime2026}. The framework could be further extended by replacing a reaction--diffusion type system with a reactive phase-separation model for a condensate-forming protein. In this case,
spatial organization would arise from the coexistence of dense and dilute phases rather than from a finite-wavelength Turing instability. Extending the
thin-channel reduction to an advective Cahn--Hilliard model would allow studying how shear flow modifies condensate
coarsening, steady droplet size, and axial localization in confined geometries. This direction is relevant to systems such as P granules in the \textit{C. elegans} zygote,  whose spatial localization is regulated through condensation and dissolution~\cite{brangwynneGermlineGranulesAre2009}, and to reactive-emulsion models in which chemical reactions control condensate size and stability~\cite{weberPhysicsActiveEmulsions2019, janssenKineticTheoryEmulsions2026,dattFluidFlowSpatiotemporal2026a}. 

\section{Acknowledgments}
We are grateful to Pierre Haas and Fridtjof Brauns for helpful discussions and insightful comments on our manuscript.

\appendix
\section{Numerical simulations}
\label{app:numerics}
We solve the full two-dimensional equations~\eqref{eq:brusselator_nd_1}--\eqref{eq:brusselator_nd_2} on a periodic axial domain $x\in[0,L_x)$ and across the channel, $y \in[-1,1]$, with no-flux walls, in the frame moving with
the mean flow, where the velocity profile is $W_x(y)-1$. 
We integrate the equations with Strang splitting~\cite{hundsdorferNumericalSolutionTimeDependent2003}, a  method in which reaction and transport are advanced separately: each time step consists of a half step of reaction, a full step of transport (advection and diffusion), and a second half step of reaction. Since transport is linear, it is solved exactly in Fourier space along the channel, with the same discretization across the channel as in Sec.~\ref{sec:numerical_linear}. The reaction is advanced with an explicit midpoint (RK2) scheme, with sub-steps where it is stiff. We use a time step $\Delta t=2\times10^{-3}$, $N_y=32$ points across the channel, and at least $128$ points along it. Runs start from the homogeneous state $(A,B/A)$ perturbed by white noise of relative amplitude $p=10^{-3}$, independent at every grid point.

From each run we keep the cross-sectionally averaged activator $\langle u\rangle (x,t)$ over the second half of the run, $t\in[T/2,T]$. We then compute two quantities: its spatial spread $s$ (standard deviation along $x$, averaged over time), which measures a pattern, the temporal spread $\tau$ of its spatial mean, which measures a uniform oscillation. A run is classified as stable if $s/A$ and  $\tau/A$ are below $10^{-3}$, i.e. if it sits at the fixed point $(A,B/A)$. Otherwise it is classified as Turing if $s/\tau>3$, as Hopf if $s/\tau<1/3$, and as undetermined if $1/3 \le s/\tau \le 3$, where neither spread dominates. The thresholds were determined empirically. The ratio $3$ was chosen from inspection of the kymographs so that a label is only assigned when one spread clearly dominates, and $10^{-3}$ is the amplitude of the initial noise, so that a run is stable only if no perturbation has grown. Phase boundaries are resolved to the grid, $\Delta A=0.1$ and $\Delta B=0.5$.

The wavelength is read off the power spectrum of $\langle u \rangle$ along $x$: it is the fundamental mode $n$, i.e.\ the lowest strong mode, so that the harmonics of a peaked pattern are not mistaken for it, and $\Lambda=L_x/n$. The box length $L_x$ is chosen between $1$ and $4$ (in units of $L$) so that it holds at least ten periods, and each run lasts long enough for the pattern to saturate.

\bibliography{bibliography}

\end{document}